# Modelling of liquid internal energy and heat capacity over a wide pressure-temperature range from first principles

J.E. Proctor

Materials and Physics Research Group, School of Science, Engineering and Environment, University of Salford, Manchester M5 4WT, United Kingdom

j.e.proctor@salford.ac.uk

## Abstract

Recently there have been significant theoretical advances in our understanding of liquids and dense supercritical fluids based on their ability to support high frequency transverse (shear) waves. Here, we have constructed a new computer model using these recent theoretical findings (the phonon theory of liquid thermodynamics), to model liquid internal energy across a wide pressure-temperature range. We have applied it to a number of real liquids in both the subcritical regime and the supercritical regime, in which the liquid state is demarcated by the Frenkel line. Our fitting to experimental data in a wide pressure-temperature range has allowed us to test the new theoretical model with hitherto unprecedented rigour. We have quantified the degree to which the prediction of internal energy and heat capacity is constrained by the different input parameters: The liquid relaxation time (initially obtained from the viscosity), the Debye wavenumber and the infinite-frequency shear modulus. The model is successfully applied to output the internal energy and heat capacity data for several different fluids (Ar, Ne, $N_2$, Kr) over a range of densities and temperatures. We find that the predicted heat capacities are extremely sensitive to the values used for the liquid relaxation time. If these are calculated directly from the viscosity data then, in some cases, changes within the margins of experimental error in the viscosity data can cause the heat capacity to exhibit a completely different trend as a function of temperature. Our code is computationally inexpensive, and it is available for other researchers to use.

1. Introduction

The heat capacity of liquids, and of supercritical fluids at liquidlike density displays – at first sight – highly unusual behaviour, since on an isochoric or isobaric temperature increase it typically decreases rather than increases. In contrast, the heat capacity of gases and solids stays constant upon temperature increase or increases, as thermal energy becomes available for a wider spectrum of excitations. The heat capacity of fluids at liquid-like density has therefore not historically been modelled from first principles. The conventional method used to model dense fluid heat capacity is the fundamental equation of state, an empirical expression for the Helmholtz free energy of the fluid $H(\rho, T)$ as a function of the density ($\rho$) and temperature ($T$) from which the heat capacities, as well as many other parameters such as the pressure $P$, can be derived[1,2,3]. The prediction of the fluid heat capacity under the conditions for which $H(\rho, T)$ was fitted to experimental data are highly accurate. However the drawbacks of this method are the mathematical and computational complexity, and the lack of relation to first principles which means that extrapolation to model $\rho, T$ conditions or mixture compositions for which the expression for the Helmholtz free energy was not fitted to experimental data is not reliable.

Thus, alternate methods to model the fluid heat capacity have been developed in recent years, in particular by accounting for the ability of dense fluids to support high frequency transverse waves. This ability has been known about for some time, having been studied with molecular dynamics simulations[4,5] and observed in a limited number of experiments with methods such as Brillouin spectroscopy[6]. What has happened only in recent years is that the existence of tranverse waves in dense fluids has been incorporated into a simple Debye-like model to predict the heat capacity of these fluids. In particular, the decrease in heat capacity upon temperature increase has been understood as being due to the spectrum of transverse waves that can be supported by the fluid decreasing as temperature rises[7,8,9]. In 2012 the heat capacity of a variety of liquids[a] was modelled (albeit over a limited range of one isobar per fluid)[7], using Eq. (1) to account for the contribution ($U_T$) of the transverse waves to the internal energy:

$$U_T(\omega > \omega_F) = 2\left(1 + \frac{1}{2}\beta T\right) \int_{\omega_F}^{\omega_D} \frac{3N_A}{\omega_D^3} \times \frac{\hbar \omega^3 d\omega}{e^{\frac{\hbar \omega}{k_B T}} - 1}$$

(1)

Here, $\beta$ is the thermal expansion coefficient and $N_A$ is Avogadro's number. This model is referred to as the phonon theory of liquid thermodynamics, and has been applied also to nanofluids and wetting effects[10,11]. Equation 1 arises from the hypothesis that transverse waves can be supported between a certain minimum frequency $\omega_F$ (calculated from Frenkel's temperature-dependent liquid relaxation time $\tau_{R(F)}(T)$) and the Debye frequency $\omega_D$. Experimental measurements of Frenkel's (atomistic) liquid relaxation time are only available for a small selection of extremely viscous liquids close to the melting curve. They are not available for typical liquids over a wide $\rho, T$ range as would be required for modelling trends in liquid heat capacity. However, $\tau_{R(F)}(T)$ can be estimated by equating it to Maxwell's (macroscopic) liquid relaxation time $\tau_{R(M)}(T)$. The Maxwell relaxation time can be obtained from the infinite-frequency shear modulus and the temperature-dependent experimentally measured viscosity[4]:

---

[a] Here, and subsequently, we use the term "liquid" to describe a fluid at supercritical temperature on the liquid-like (high density) side of the Frenkel line, as well as at subcritical temperature.

$$\tau_{R(M)}(T) = \frac{\eta(T)}{G_\infty}$$

(2)

Thus the internal energy (and hence the heat capacity) can be modelled using the experimental data for the viscosity and thermal expansion coefficient along with just two adjustable parameters: $\omega_D$ and $G_\infty$ (the infinite frequency shear modulus). These took physically reasonable values for the liquids studied in ref. 7.

There is, however, much development and testing of the phonon theory of liquid thermodynamics still required. The present work tackles four areas in which further work is needed:

Firstly, the manner in which it is necessary to use empirically adjustable parameters to model the heat capacity is not clear at present. Any model for liquid heat capacity requires the use of empirically adjustable parameters since experimental measurements are not widely available of all the parameters required; So the use of adjustable parameters does not necessarily indicate a shortcoming in the model. In ref. 7 the Debye frequency was set to the value observed for the solid state of each substance studied, then a physically reasonable value of $G_\infty$ was obtained by a manual parameter fitting process. It is thus not clear if the values chosen are those that provide the best fit to the data. What are the consequences if values which are not physically reasonable turn out to provide a better fit to the data?

Secondly, to date no testing has taken place in which the theory has been used to predict the heat capacity along multiple $P, T$ or $\rho, T$ paths for the same fluid. What are the consequences if the best-fit values of the fitting parameters do not exhibit the expected trends when compared across different paths? For instance, if several $\rho, T$ paths are examined the best-fit value of $G_\infty$ should not decrease upon density increase. What if it does?

Thirdly, by definition $C_V \equiv (\partial U/\partial T)_V$ but implementations of the phonon theory of liquid thermodynamics to date have focussed on studying $U$ along isobars. In this case a direct differentiation of $U$ provides $(\partial U/\partial T)_P$. It is not clear what additional steps have been taken to obtain $(\partial U/\partial T)_V$ from the data along isobars.

Fourthly, a more detailed look at the viscosity data is required. Thus far, the viscosity data required to calculate $\tau_{R(M)}(T)$ has been fitted with the VFT (Vogel-Fulcher-Tammann) law (given later in Eq. (3)) before input into Eq. (2)[12]. The VFT law has some theoretical foundation at low temperatures[13] but it is not clear if it provides a good fit to the data throughout the wide $P, T$ range in which we would like to calculate $\tau_{R(M)}(T)$ from the viscosity, whilst retaining physically realistic values of the parameters (particularly $T_0 > 0$). In some preliminary work, we found that the obtained $C_V$ could be drastically affected by small changes in the input values used for the viscosity, as has been noted elsewhere[12]. It is thus crucial that the manner in which the viscosity data has been prepared for input into Eq. (2) is laid out in a clear and unambiguous manner.

In this work, we present an improved and computationally inexpensive methodology for modelling the heat capacity of a variety of liquids over a wide $\rho, T$ range including at significantly supercritical temperature on the liquid side of the Frenkel line and at the lowest temperatures where the liquid phase exists, close to the triple point. We provide an explicit clarification of what parameters are empirically adjustable and how they were adjusted, and provide full details of the methodology to enable our work to be applied by other researchers. This can provide a firm foundation for future work in which this and similar models are applied to the heat capacity on the transition across the Frenkel line into the gas state, and to $\rho, T$ conditions and fluid mixture compositions for which experimental data are not yet available.

## 2. Development of the fluid energy model

In this section we outline the physical principles underpinning the development of our "fluid energy" computer model to fit observed liquid heat capacities from first principles. The model is implemented in C++ using our own code to search the parameter space and perform nonlinear regression analysis, and using code from the Gnu Scientific Library[14] to perform numerical integration for calculations equivalent to Eq. (1). We have used our model to model the heat capacity along isochores since parameters such as $k_D, G_\infty$ are likely to vary less along isochores than along isobars, and because a direct differentiation of the internal energy provides the heat capacity $C_V = (\partial U/\partial T)_V$ without any additional steps. Minor computational details are dealt with in the supplementary information.

### 2.1 Fitting to the output from the fundamental equation of state

To begin, it is appropriate to outline certain matters relating to the experimental data with which we are comparing the output of our model. We compare our output to the output provided by the relevant fundamental equation of state available via NIST REFPROP[15]. The fundamental equation of state is a mathematical model giving the Helmholtz free energy of the fluid (in the liquid, supercritical fluid and gas states) as a function of density and temperature: $H(\rho, T)$. It is not possible to directly measure this quantity experimentally but a wide variety of experimentally measurable properties of the fluid (for instance pressure, heat capacities, speed of longitudinal sound waves) can be obtained from $H(\rho, T)$ using mathematical relations obtained directly from first principles. The fundamental equation for $H(\rho, T)$ for a specific fluid is a complex equation containing ca. 50 empirically adjustable parameters, and many terms (the "bank of terms" in the residual part) in which the mathematical forms of the terms are chosen empirically, not just the parameter values[1-3]. The adjustable parameters are obtained using nonlinear regression analysis to result in an expression for $H(\rho, T)$ which accurately models the available experimental data. It is common for the fundamental EOS to model all available data to an accuracy of $\pm \sim 0.5\%$ so here, and elsewhere, the output from the fundamental EOS is simply referred to as "data" and is treated as experimental data. However it is not a model obtained from first principles so cannot be reliably extrapolated to predict fluid properties under conditions where it has not been fitted, or at least compared, to the original experimental data.

For this work, we utilize the fundamental equation of state outputs for the internal energy $U$, longitudinal sound speed $c_L$, viscosity $\eta$ and pressure $P$. All of these parameters can be obtained directly from $H(\rho, T)$ via NIST REFPROP and further details found in the relevant publications[16-19]. In the earlier work[7,12] to model fluid heat capacity from first principles using the phonon theory of liquid thermodynamics, an expression for the internal energy incorporating Eq. (1) and an equivalent equation for the contribution from longitudinal waves was numerically differentiated and the results manually compared to the values of $C_V$ obtained from $H(\rho, T)$ via NIST REFPROP.

Obtaining the best values of any adjustable parameters using a proper nonlinear regression analysis procedure fitting to $C_V, C_P$ directly would be challenging since these depend on derivatives of the quantity we are calculating (for instance Eq. (1) for $U_T$ and an equivalent expression for $U_L$).

Therefore in our model we fit to the internal energy $U$ directly and then compare our results to the observed heat capacities. This enables us to use nonlinear regression analysis to obtain the best values of adjustable parameters in our work. Whilst it is not guaranteed that nonlinear regression analysis will provide the best fit (only linear regression analysis can do that), it is the best approach available.

### 2.2 Preparation of viscosity data

The Frenkel relaxation time $\tau_{R(F)}$ is not directly measurable; it is defined atomistically as the length of time a fluid particle spends in an equilibrium position before moving to a new equilibrium position. However, it is expected to be the same order of magnitude as the Maxwell relaxation time $\tau_{R(M)}$ which is defined using Eq. (2) in terms of macroscopic fluid properties. Generally, experimental measurements of $G_\infty$ are not available but it is not expected to vary significantly as a function of temperature at constant pressure or density. We therefore assume that it is independent of temperature along an isochore and treat it as a single fitting parameter. The viscosity of both liquids and gases varies with temperature. In a liquid at subcritical temperature the viscosity roughly follows the VFT law[13] (Eq. (3)) in which the viscosity decreases upon temperature increase, although other theoretical models are available[20,21].

$$\eta(T) = \eta_\infty e^{A/(T-T_0)}$$

(3)

In the gas state viscosity is independent of density and follows the simple relation $\eta(T) = J\sqrt{T}$, increasing as a function of temperature[22] contrary to the trend observed in liquids. Here, we wish to model the heat capacity of liquids at supercritical temperature on the liquid side of the Frenkel line. Under these conditions the viscosity, whilst not following the simple $\sqrt{T}$ dependence observed for gases, does begin to increase upon (isochoric) temperature increase. The raw viscosity data cannot therefore be used to calculate the relaxation time according to Eq. (2) as the viscosity is transitioning between liquid-like and gas-like behaviour. Our solution is to fit the viscosity data using Eq. (4), separating out the components representing liquid-like and gas-like behaviour. This regression analysis is performed using Magicplot Pro and the parameter $J$ is supplied to the fluid energy program along with the raw viscosity data so that the component representing gas-like behaviour can be subtracted from the viscosity data before use. The only constraint in the regression analysis procedure is that $T_0 > 0$ as it is only physically realistic for $T_0$ to take positive values. In some cases we find that $T_0 = 0$ provides the best fit, in which case Eq. (3) simplifies to the Arrhenius equation. Figure 1 shows an example of this fit, the Ar (argon) isochore from 120 K – 500 K at 30.0 mol L$^{-1}$.

$$\eta(T) = \eta_\infty e^{A/(T-T_0)} + J\sqrt{T}$$

(4)

Henceforth the liquid-like viscosity component that is used by the fluid energy program will be denoted by $\eta_L$, where $\eta_L = \eta - J\sqrt{T}$. Thus the only processing of the viscosity data is to subtract the component representing gas-like behaviour, modelled using the $J\sqrt{T}$ term which is the temperature dependence of viscosity in a gas expected from first principles. The only way in which the overall fit to the viscosity data from Eq. (4) is utilized is to select the value of $J$.

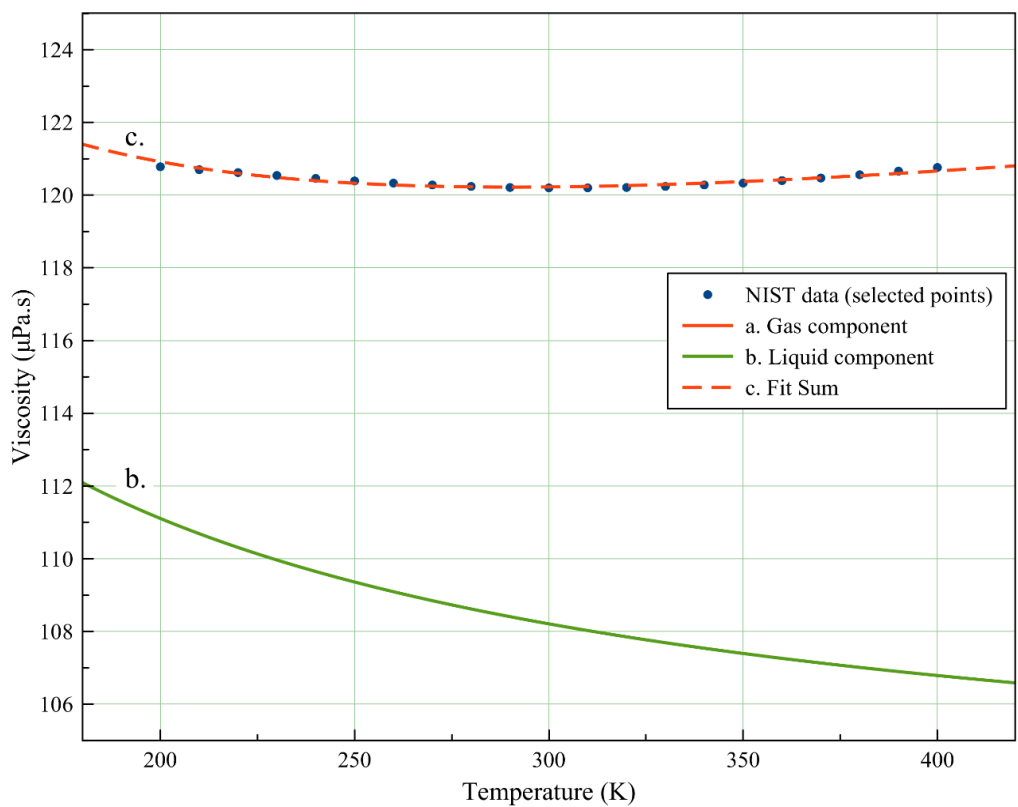

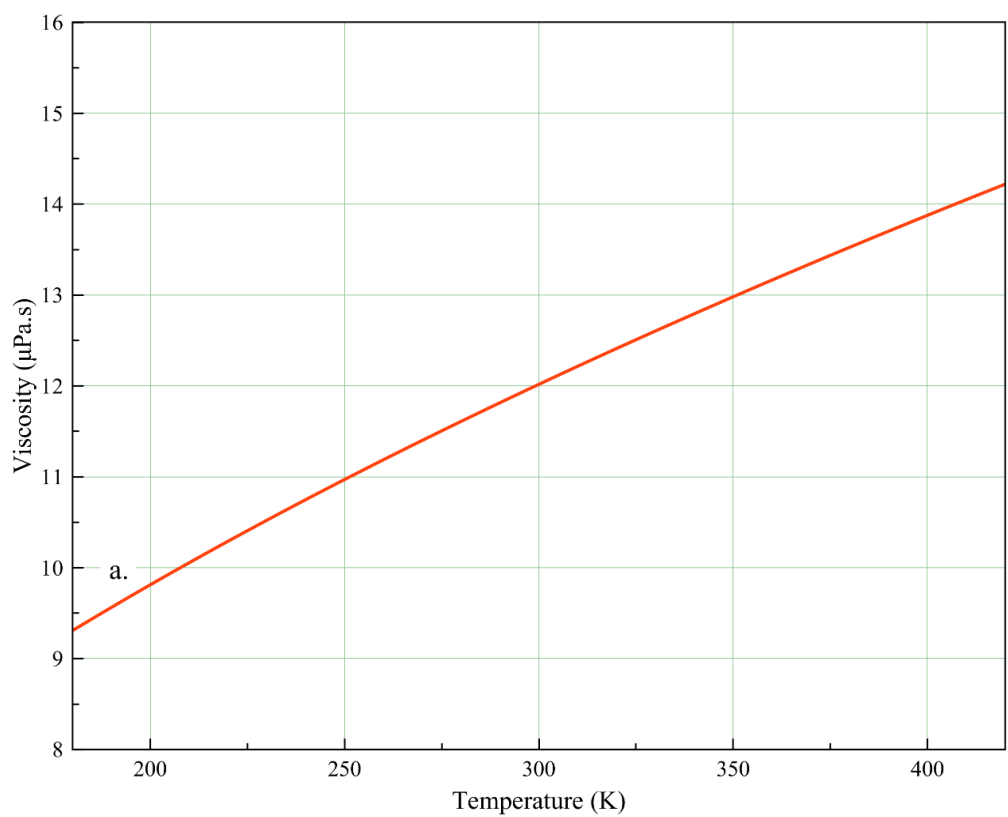

Figure 1. Viscosity of fluid Ar at 30.0 Mol.L$^{-1}$ density. Selected experimental values from NIST are shown along with the fit components utilizing Eq. (4). The gas component (red line) corresponds to $J = 0.69 \pm 0.02$ µPa.s.K$^{-1/2}$, where the error is obtained from the regression analysis.

### 2.3 k-gap versus frequency gap

Eq. (1) arises directly from Frenkel's proposal of the liquid relaxation time $\tau_{R(F)}$. The dispersion relation for transverse waves is assumed to remain linear, but waves with longer period than $\tau_{R(F)}$ cannot propagate. There is therefore a "gap" in both the frequency / energy domain and the wavenumber domain of missing transverse waves expressed in the non-zero lower integral limit in Eq. (1). Since then, it has been proposed that the gap in fact exists in the wavenumber domain, the so-called "k-gap"[23,24]. In this case, the dispersion relation for transverse waves is given by:

$$\omega_T(k) = \sqrt{c_T^2 k^2 - \frac{1}{4\tau_{R(F)}^2}}$$

(5)

This dispersion relation is obtained by solving a form of the Navier-Stokes equation, modified to consider the effect of short-term elasticity in the fluid. Equation (5) has also been given utilizing $\tau_{R(M)}$ in place of $\tau_{R(F)}$[4]. The questions of how similar these relaxation times are, and which is the correct one to use in Eq. (5), are the subject of ongoing debate[25,26]. In either case $\omega_T(k_{gap}) = 0$ where $k_{gap} = 1/2c_T\tau_{R(F/M)}$ so modes with $k < k_{gap}$ do not exist. Modes with $k_{gap} < k < k_{gap}\sqrt{2}$ are overdamped so do not propagate[23]. For $k > k_{gap}\sqrt{2}$ the frequencies obtained using Eq. (5) and $\omega_T(k) = c_T k$ differ only by a factor of $\leq \sqrt{2}$. The discrepancy decreases on further increase of $\omega, k$ and due to the increased energy of each individual excitation combined with the increasing density of states ($\propto k^2$) these are the modes which have the dominant effect on $U$.

It is probably as a result of these factors that we have not found it necessary to incorporate the non-linear dispersion relation (5) into our model and we have made only one change resulting from the k-gap theory. In ref. 7 the total internal energy is given as: $U = U_L + U_T(\omega > \omega_F) + U_T(\omega < \omega_F)/2$. Here $U_L$ is the contribution from the longitudinal modes. The final term in this expression accounts for the fact that, upon temperature increase, whilst the potential energy contribution from transverse modes is lost, the kinetic energy contribution remains; becoming the translational degrees of freedom that exist also in the gas state. The kinetic energy contribution is 1/2 of the total energy, a result obtained from the virial theorem.

Using the k-gap model, it is no longer appropriate to account for this using a term of the form $U_T(k < k_{prop})/2$ as with the dispersion relation given by Eq. (5) the low-k modes do not have a real energy. We therefore use:

$$U = U_L + \frac{U_T(k > k_p)}{2} + RT$$

(6)

Here, the $U_T/2$ term accounts for the potential energy associated with the transverse wave degrees of freedom (which is gradually lost upon temperature increase) and the $RT$ term for the kinetic energy associated with these degrees of freedom.

### 2.4 The fluid energy model for $U$

The expressions used here for the contributions to the fluid internal energy are similar to those in previous works[7] but the derivations are given here to clarify what assumptions / approximations are made in the derivation. The objective is to evaluate the contributions to the fluid internal energy arising from longitudinal and transverse collective vibrational modes that can be treated using methods analogous to phonons in solids. We begin with the expression from Landau and Lifshitz[27] for the contribution ($F_{ph}$) to the Helmholtz free energy of a solid arising from these vibrational modes:

$$F_{ph} = k_B T \sum_i \ln\left[1 - e^{-\frac{\hbar \omega_i}{k_B T}}\right]$$

The internal energy contribution $U_{ph}$ is obtained by differentiating at constant volume[27,28]:

$$U_{ph} = F_{ph} - T\left(\frac{\partial F_{ph}}{\partial T}\right)_V$$

Performing this differentiation exactly as given we obtain:

$$U_{ph} = \sum_i \frac{1}{e^{\frac{\hbar \omega_i}{k_B T}} - 1}\left[\hbar \omega_i - \hbar T \left(\frac{\partial \omega_i}{\partial T}\right)_V\right]$$

There are two approaches to deal with the remaining partial derivative. The first is to assume that The Grüneisen approximation holds exactly as written, which is that both the temperature and pressure dependence of the vibrational frequencies can be incorporated into the volume dependence. In this case,

$$\frac{\omega_i}{\omega_i^0} = \left[\frac{V(T,P)}{V_0}\right]^{-\gamma_i}$$

$$\left(\frac{\partial \omega_i}{\partial T}\right)_V = 0, \quad \forall\, i$$

Where $\gamma_i$ is the Grüneisen parameter for mode $i$. Alternatively the following law has been proposed to hold at liquid-like density[7,29]:

$$\left(\frac{\partial \omega_i}{\partial T}\right)_V = -\frac{1}{2}\beta \omega_i$$

Where $\beta$ is the thermal expansion coefficient. We thus introduce the parameter λ which is set to zero if the Grüneisen approximation is assumed to hold, and 0.5 if the equation above is used. The final expression for the internal energy is therefore:

$$U_{ph} = (1 + \lambda \beta T) \sum_i \frac{\hbar \omega_i}{e^{\frac{\hbar \omega_i}{k_B T}} - 1}$$

The components in our final expression for $U$ (equation (6)) are obtained from the above expression and given in Eq. (7) below. The sum has been replaced by an integral and the density of states $g(k)$ (normalized for the correct number of modes to exist up to $k_D$) has been introduced, for a sample comprising one mole of particles.

$$U_L = (1 + \lambda \beta(T) T) \int_0^{k_D} g(k) \frac{\hbar \omega_L(k) dk}{e^{\frac{\hbar \omega_L(k)}{k_B T}} - 1}$$

$$U_T = 2(1 + \lambda\beta(T)T) \int_{k_{prop}}^{k_D} g(k) \frac{\hbar\omega_T(k)dk}{e^{\frac{\hbar\omega_T(k)}{k_B T}} - 1}$$

$$g(k) = 3N_A k^2 / k_D^3$$

$$\omega_L(k) = c_L k$$

$$\omega_T(k) = c_T k$$

(7)

The factor of 2 in the expression for $U_T$ is to account for the 2 degrees of freedom existing perpendicular to the propagation axis. The parameter $c_L$ is the speed of longitudinal sound waves. We use the experimental data from NIST with no intermediate approximations / fitting. The parameter $c_T$ is the speed of the transverse sound waves. There are far fewer data on this available but it can be calculated from $G_\infty, \rho$ using Eq. (8):

$$c_T = \sqrt{\frac{G_\infty}{\rho_{mass}}}$$

(8)

Here, $\rho_{mass}$ is the constant mass density along the isochore, and $G_\infty$ is the fitting parameter utilized also in the calculation of $\tau_{R(M)}$. $G_\infty$ is assumed to stay constant along the isochore. The lower wavenumber limit $k_{prop}(M)$ for transverse waves is calculated from $k_{gap}$ obtained from Eq. (5), utilizing Eqs. (2) and (8).

$$k_{prop}(M) = \frac{\sqrt{\rho_{mass} G_\infty}}{\eta_L \sqrt{2}}$$

(9)

The label $(M)$ indicates that the parameter has been calculated directly from Maxwell's liquid relaxation time as given in Eq. (2). The thermal expansion coefficient $\beta(T)$ is obtained directly from the experimental data at each temperature. The different approaches to the parameter $\lambda$ were outlined earlier. Since the primary purpose of the present work is to test theoretical proposals against experimental data, the fluid energy model compares the fit to experimental data obtained with the values $\lambda = 0, 0.5$. The input parameters used to model the fluid internal energy and heat capacity along an isochore are summarized in table 1.

The principal fitting parameters utilized by the fluid energy program are hence $k_D$ and $G_\infty$, whilst $\lambda$ can also be fitted to some extent (values of 0 and 0.5 are compared). Therefore, in addition to producing the output of $U$ as a function of temperature for the best fit values of these parameters, the program produces (for the best fit values of $k_D, \lambda$) the overall least-squares error in $U$ as a function of $G_\infty$, and the equivalent data for $k_D$. It is therefore possible to evaluate how strongly the fit quality varies as a function of these variables.

| Parameter | Source |
|---|---|
| Thermal expansion coefficient $\beta(T)$ | Calculated from NIST data at each temperature on the isochore. |
| Anharmonicity parameter $\lambda$ | Fits for $\lambda = 0$, $\lambda = 0.5$ are compared. |
| Mass density $\rho_{mass}$ | Constant along each isochore. |
| Debye wavenumber $k_D$ | Fitting parameter obtained by the fluid energy program using our own nonlinear regression analysis. |
| Infinite-frequency shear modulus $G_\infty$ | Fitting parameter obtained by the fluid energy program using our own nonlinear regression analysis. |
| Viscosity $\eta$ | Obtained from NIST data. |
| Gas-like viscosity coefficient $J$ | Obtained using nonlinear regression analysis in Magicplot Pro. |

Table 1. Inputs used by the fluid energy program.

### 2.5 Testing on liquid and supercritical Ar

We initially tested the fluid energy program by studying fluid Ar along two isochores (30 Mol L$^{-1}$ and 32.5 Mol L$^{-1}$) covering temperature range beginning in the subcritical region then extending as far as the Frenkel line in the supercritical region. Figure 2 is the $\rho, T$ phase diagram of fluid Ar (further details given in ref. 30), on which these isochores are illustrated.

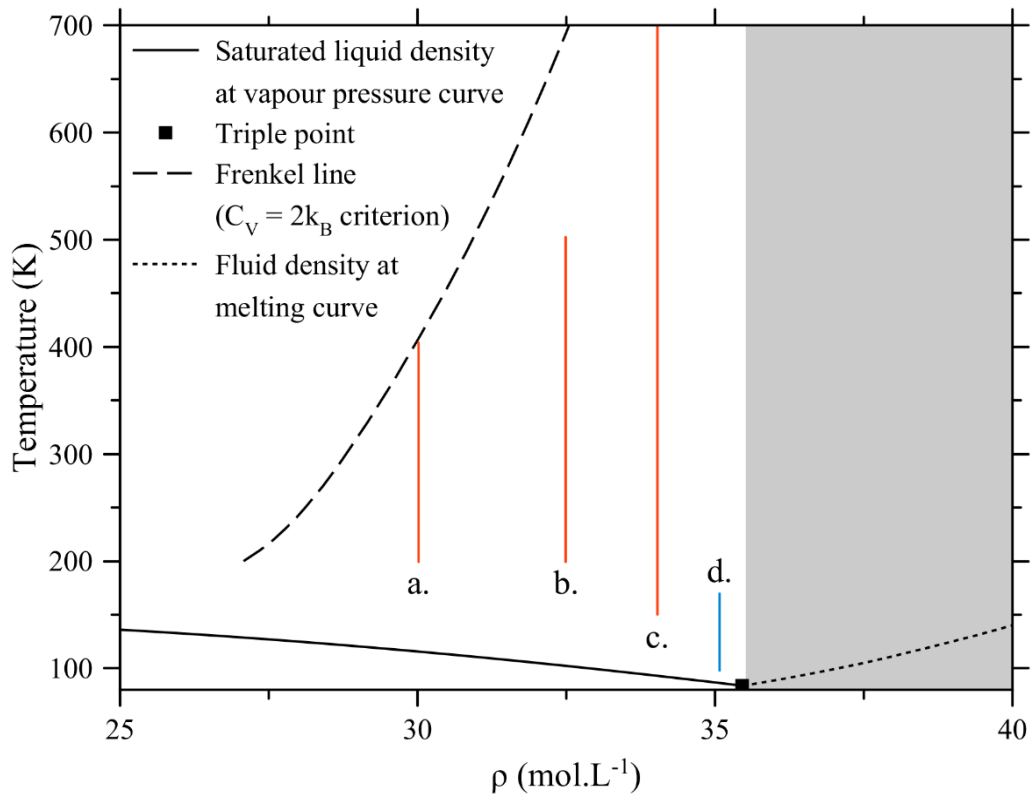

Figure 2. $\rho, T$ phase diagram of fluid Ar. Red lines (a, b and c) denote the isochores studied in the development of the fluid energy program and the blue line (d) denotes the isochore studied during our investigations of the subcritical liquid state in section 3.3. The grey shaded area indicates the region not covered by the fundamental equation of state data.

Figure 3 shows the results of a systematic search of the $\lambda, k_D, G_\infty$ parameter space on the 30 Mol.L$^{-1}$ isochore. In this run, 2 x 50$^2$ points in the parameter space were searched: 50 values of $k_D$ (1.0 – 1.4 x 10$^{10}$ m$^{-1}$) combined with 50 values of $G_\infty$ (0.1 – 5.0 GPa) and $\lambda = 0, 0.5$. The values of $k_D$ were chosen to be close to the value of 1.2 x 10$^{10}$ m$^{-1}$ estimated from comparison to the solid state (see supplementary information) and values of $G_\infty$ were selected to cover a wide range. As shown in the figure, and in similar results obtained at 32.5 Mol.L$^{-1}$, a reasonable fit to the internal energy can be obtained using the methodology described thus far. However, the heat capacities are obtained from the derivative $(\partial U/\partial T)_V$ so it is necessary to obtain a much better fit to the internal energy to obtain a good prediction of the heat capacities. The next section describes how this is achieved using the last stage of regression analysis in the fluid energy program.

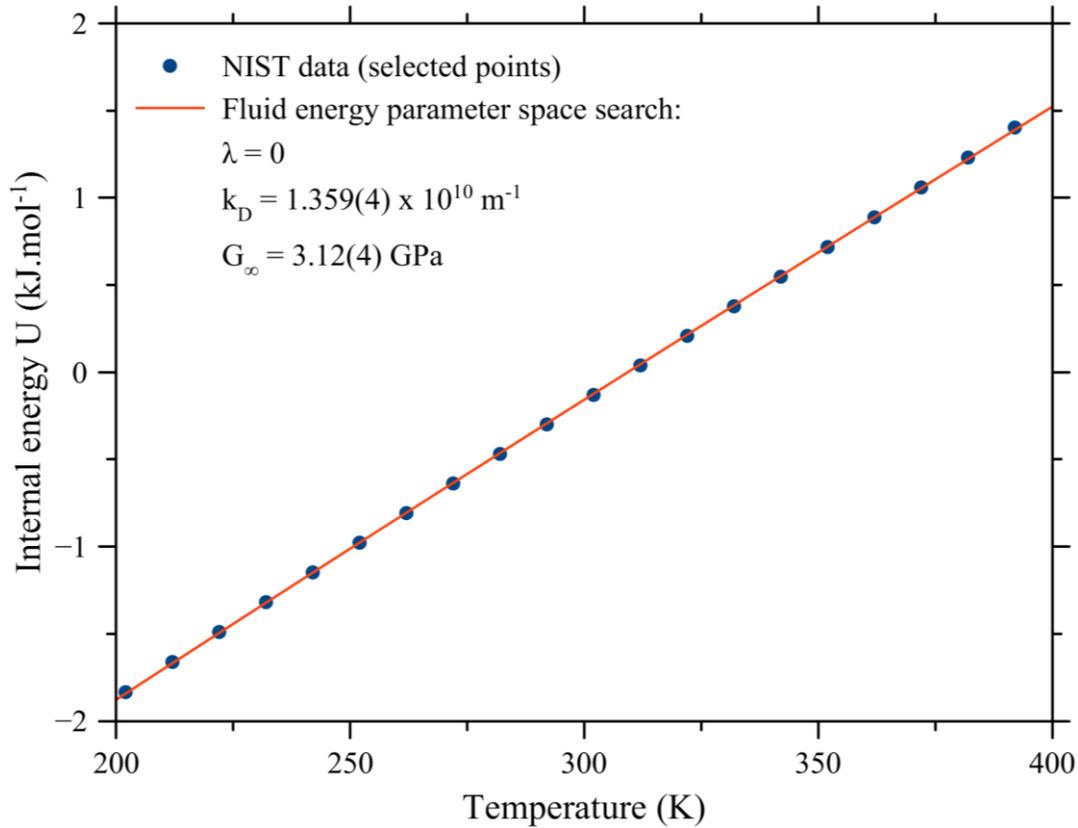

Figure 3. Results of 2x50x50 parameter space search to fit to the experimentally measured internal energy of fluid Ar at 30 Mol.L$^{-1}$. Parameter values shown are those resulting in the best fit to the data.

### 2.6 Refinement of the minimum shear wavenumber $k_{prop}$

The parameter $k_{prop}$ appearing as the lower limit to the integral for $U_T$ in Eq. (7) has been calculated thus far using Eq. (9) which is derived from the expression for Maxwell relaxation time $\tau_{R(M)}$. However, it has also been proposed to calculate it using the Frenkel relaxation time $\tau_{R(F)}$[23,24]. Whilst these relaxation times are expected to be the same order of magnitude, they are not the same quantity. The last stage in the fitting process in the fluid energy program allows this degree of freedom by refining $k_{prop}$ at each temperature to obtain a value for $U$ that most closely matches the NIST data. Figure 4 (a) compares the values of $k_{prop}$ obtained using Eq. (9) ($k_{prop}(M)$) and the values taken by $k_{prop}$ following the refinement process ($k_{prop}(F)$) with the 30 Mol.L$^{-1}$ Ar data shown in figure 3. Figure 4 (b) shows the results for $C_V$ obtained following the parameter search and refinement processes. As shown, the parameter search process produces a poor fit for $C_V$. This problem is rectified by the refinement process, despite the modest changes to $k_{prop}$ (< 0.3% at 30 Mol.L$^{-1}$ and < 1.3% at 32.5 Mol.L$^{-1}$). The nonlinear regression analysis code utilized for the refinement is described in the supplementary information.

Note that the freedom given to $k_{prop}$ here is greater than the freedom given at the earlier stage (searching the $G_\infty, k_D$ parameter space). At the earlier stage the absolute value of $k_{prop}$ could be varied by changing $G_\infty$, but since $G_\infty$ is independent of temperature in our model the variation of $k_{prop}$ as a function of temperature was determined solely by the temperature-dependent viscosity $\eta_L$. At this stage however, we also allow the temperature dependence of $k_{prop}$ to be different from the temperature dependence of $\eta_L$.

It is this part of the fitting process for which the decision to fit to $U$, rather than directly fit to $C_V$, is essential. Adjusting $k_{prop}$ at a certain temperature would cause opposite effects on the gradient $(\partial U/\partial T)_V$ on each side of the chosen temperature so getting the process to converge towards a better overall fit to $C_V$ would be very challenging.

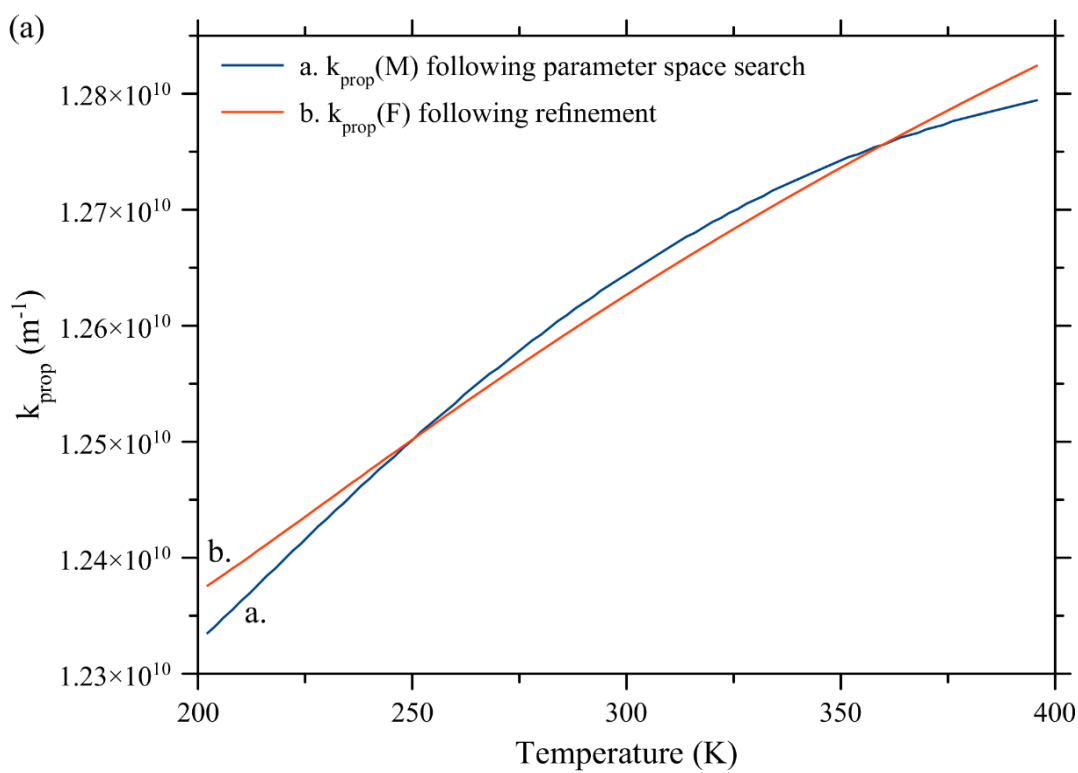

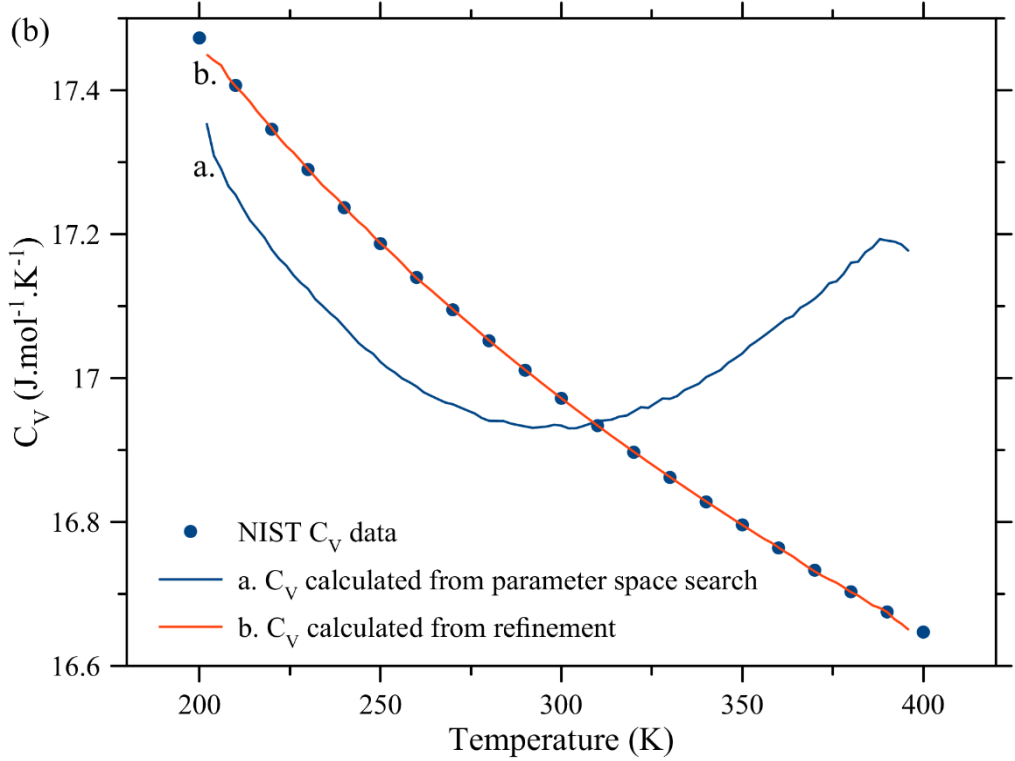

Figure 4 (a) values $k_{prop}$ for 30 Mol.L$^{-1}$ Ar following the parameter space search and refinement processes described in the text. (b) $C_V$ calculated using fluid energy program after parameter space search and refinement processes, compared to NIST data. $C_V$ data have been smoothed following numerical differentiation of $U$.

### 2.6 Validity of the fluid energy model

The fluid energy model as presented here does not remain valid if the Frenkel line is crossed into the gas / gas-like state, for two reasons. Firstly, as $k_{prop} \to k_D$ the value of the integral for $U_T$ in equation (9) approaches zero. Then, when $k_{prop} > k_D$ the integral has to be manually removed from the calculation to avoid the existence of a negative contribution to the internal energy. There would thus be a discontinuity in the gradient $(\partial U/\partial T)_V$ at $k_{prop} = k_D$. Secondly, in the gas state the contribution to $U$ from potential energy associated with longitudinal phonon-like vibrations gradually disappears also. Eq. (9) does not account for this effect. The heat capacity of gases close to the Frenkel line has been modelled in separate work[31].

The model does, however, give the correct behaviour up to the Frenkel line at $k_{prop} = k_D$. At this point $U_T = 0$ and $U = U_L + RT$ (c.f. Eq. (6)). Provided that adequate thermal energy is available to excite all longitudinal modes ($k_B T \gg \hbar c_L k_L$), Eq. (7) simplifies to $U_L \approx RT$ giving $C_V = 2R$ as expected.

In addition, the model can fail if we attempt to cover a very large temperature range in a single isochore. For instance, in the case of Ar the isochore shown in figure 2 at 34 MolL$^{-1}$, running from 150 K to 700 K (the highest temperature at which data is available from the fundamental EOS) is the largest temperature range for which a physically realistic output can be achieved. In this case (shown in figure 5) $k_{prop}$ began to show the physically unrealistic behaviour of decreasing on temperature increase at the highest and lowest temperatures in the isochore.

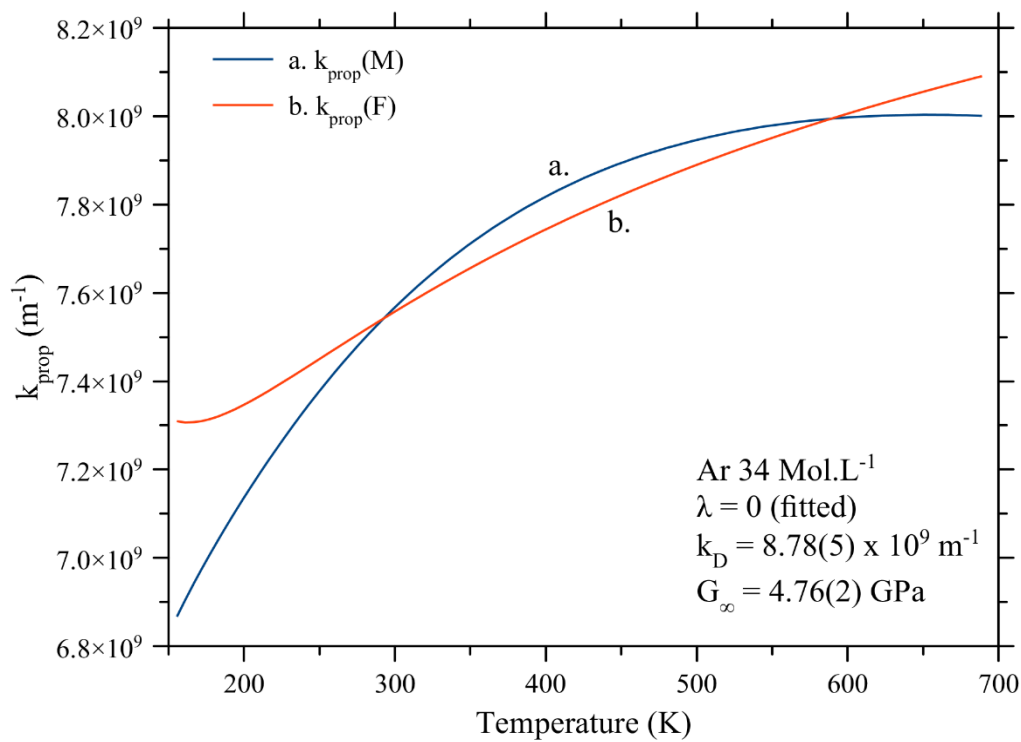

Figure 5.  Results for $k_{prop}$ following the parameter space search ($k_{prop}(M)$) and refinement ($k_{prop}(F)$) processes for Ar at 34 Mol.L⁻¹, 150 – 700 K.

## 3 Results

We selected Ne (neon) and $N_2$ (nitrogen) as suitable fluids with which to run the fluid energy program over a range of isochores in the supercritical state.  Both are simple fluids in which the Frenkel line has been characterized in some detail[32,33].  We also examined subcritical isochores close to the triple point for Ar and Kr (Krypton).

### 3.1 Ne

Internal energy and heat capacity data for fluid Ne are available from NIST up to a density of 62 Mol.L⁻¹.  We ran the fluid energy model along four isochores in the rigid liquid state, as illustrated in figure 6 (a) on the $\rho, T$ phase diagram (further details given in ref. 30).  In all cases we were able to obtain a good fit for $C_V$ with very limited changes to $k_{prop}$ during the refinement process (< 0.2%).  Since $k_{prop} \propto 1/\eta$, the experimental error in the viscosity data propagates through to $k_{prop}$.  The changes to $k_{prop}$ during the refinement process are within the error in the viscosity data, yet cause significant changes to $C_V$ which are large compared to the experimental error (5%) on this parameter.  Figure 6 (b) shows, at 60 Mol.L⁻¹, the effect on $C_V$ of the $\pm 0.2\%$ changes to $k_{prop}$ made during the refinement process.

The best-fit value of $\lambda$ was 0.5 in all four cases.  The best-fit values of $k_D, G_\infty$ took physically reasonable values, however they varied in a haphazard manner on density increase rather than displaying the trend which should exist (a small increase upon density increase in both cases).  We therefore ran the fluid energy program with $k_D = 1.42 \times 10^{10}$ m⁻¹, the average value obtained from the calculations in which $k_D$ was allowed to vary.  In this case $G_\infty$ exhibited the expected trend (a small increase on density increase). The changes to $k_{prop}$ required during the refinement process increased slightly but remained very small (<0.8%). Table 1 gives all fit parameters from the calculations on Ne.

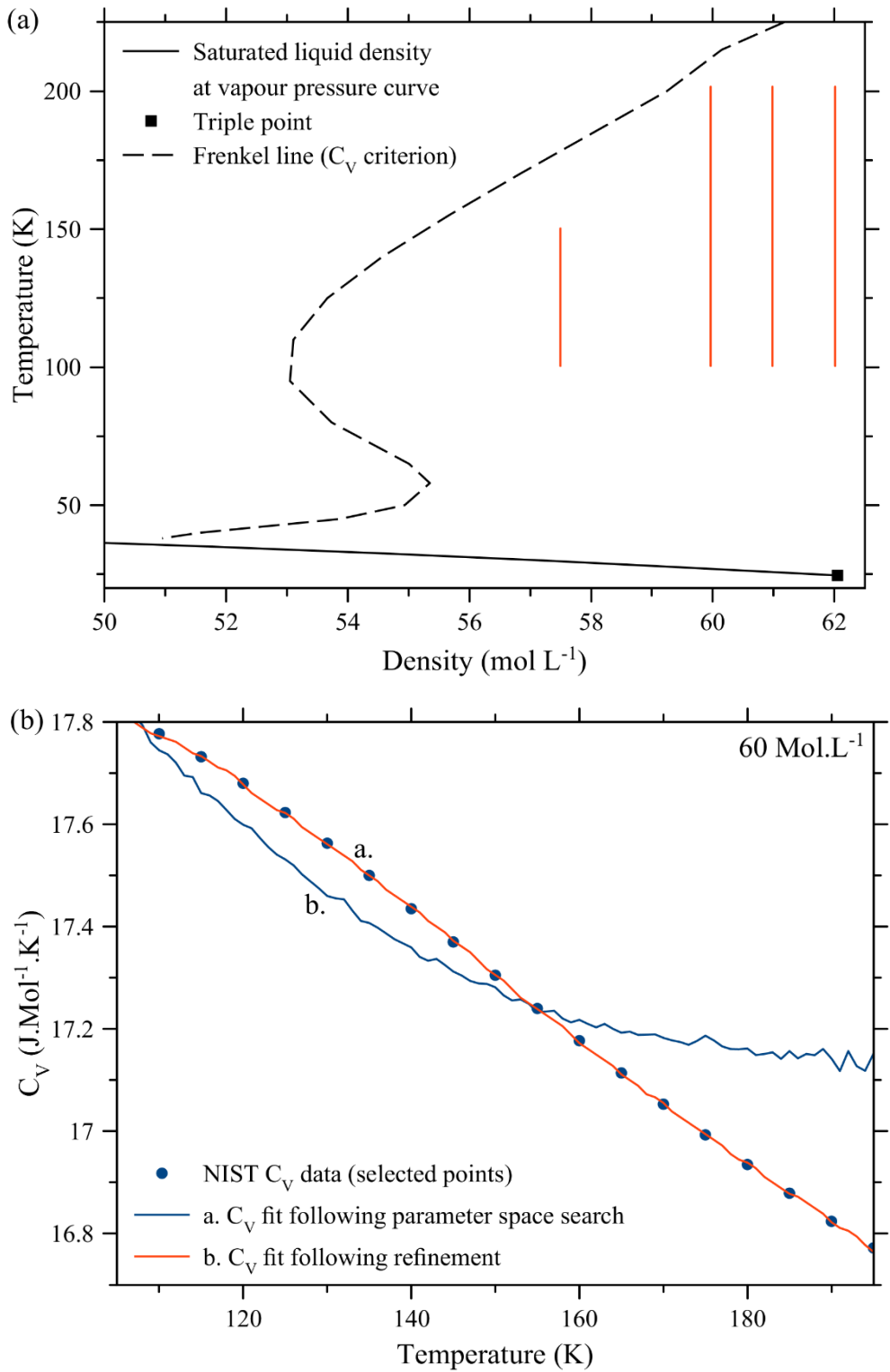

Figure 6. (a) $\rho, T$ phase diagram of fluid Ne with the isochores studied here marked. (b) Experimental $C_V$ data for fluid Ne at 60 Mol.L$^{-1}$ from NIST plotted alongside the theoretical prediction following the parameter space search and following refinement of $k_{prop}$ by up to $\pm 0.2\%$. Data have been smoothed following numerical differentiation of $U$.

| Density (Mol.L$^{-1}$) | $\lambda$ | $k_D$ (m$^{-1}$) | $G_\infty$ (GPa) |
|---|---|---|---|
| 57.5 | 0.5 (fitted) | $1.47(1) \times 10^{10}$ (fitted) | 0.89(1) (fitted) |
| 60 | 0.5 (fitted) | $1.55(1) \times 10^{10}$ (fitted) | 1.11(1) (fitted) |
| 61 | 0.5 (fitted) | $1.35(1) \times 10^{10}$ (fitted) | 0.91(1) (fitted) |
| 62 | 0.5 (fitted) | $1.31(1) \times 10^{10}$ (fitted) | 0.93(1) (fitted) |
| 57.5 | 0.0 (fitted) | $1.42 \times 10^{10}$ (fixed) | 0.74(1) (fitted) |
| 60 | 0.0 (fitted) | $1.42 \times 10^{10}$ (fixed) | 0.82(1) (fitted) |
| 61 | 0.0 (fixed) | $1.42 \times 10^{10}$ (fixed) | 0.84(1) (fitted) |
| 62 | 0.0 (fitted) | $1.42 \times 10^{10}$ (fixed) | 0.96(1) (fitted) |

Table 1. Fitted values of $k_D, G_\infty$ obtained in fluid Ne simulations described in the text.

The upper temperature bound of our investigation of fluid Ne is formed by the Frenkel line. The lower temperature bound was determined by the need to avoid the well-documented anomalies in the output from the fundamental EOS in the subcritical region[34]. Relative to the critical temperature (44.5 K), this has allowed us to study the range $T_R = T/T_C = 2 - 4$.

**3.2 N$_2$**

In our study of fluid N$_2$, it was possible to examine a greater range in absolute temperature, and also in density. The upper (315 K) and lower (160 K) temperature bounds of our study of fluid nitrogen were dictated by the need to subtract the heat capacity contribution due to the intra-molecular degrees of freedom. To check this, we examined the heat capacity data in the gas state (the 0.01 MPa isobar and 0.05 Mol.L$^{-1}$ isochore). Within this temperature range, $C_V$ stays constant at $5R/2$ to within ±0.07% as the rotational degree of freedom is excited whilst the vibrational degree of freedom is not. This was accounted for by subtracting $RT$ from the values of $U$ obtained from NIST prior to entry into the fluid energy program. Above 315 K the vibrational degree of freedom becomes excited causing a gradual change in $C_V$. Below 160 K the rotational degree of freedom must be gradually lost but it is not possible to discern at exactly what temperature this process begins as $C_V$ in the gas state rises rapidly as the sublimation curve is approached. Nevertheless, our investigation of fluid nitrogen covers $T_R = 1.27 - 2.5$.

In terms of density, our investigation runs from 25 Mol.L$^{-1}$ (the lowest density at which we can cover a significant temperature range without crossing the Frenkel line into the gas state) to 30.9 Mol.L$^{-1}$ (the highest density for which experimental heat capacity data are available). In figure 7 the ρ,T phase diagram of N$_2$ is shown with the isochores studied here included. Further details on the phase diagram data are given in ref. 30.

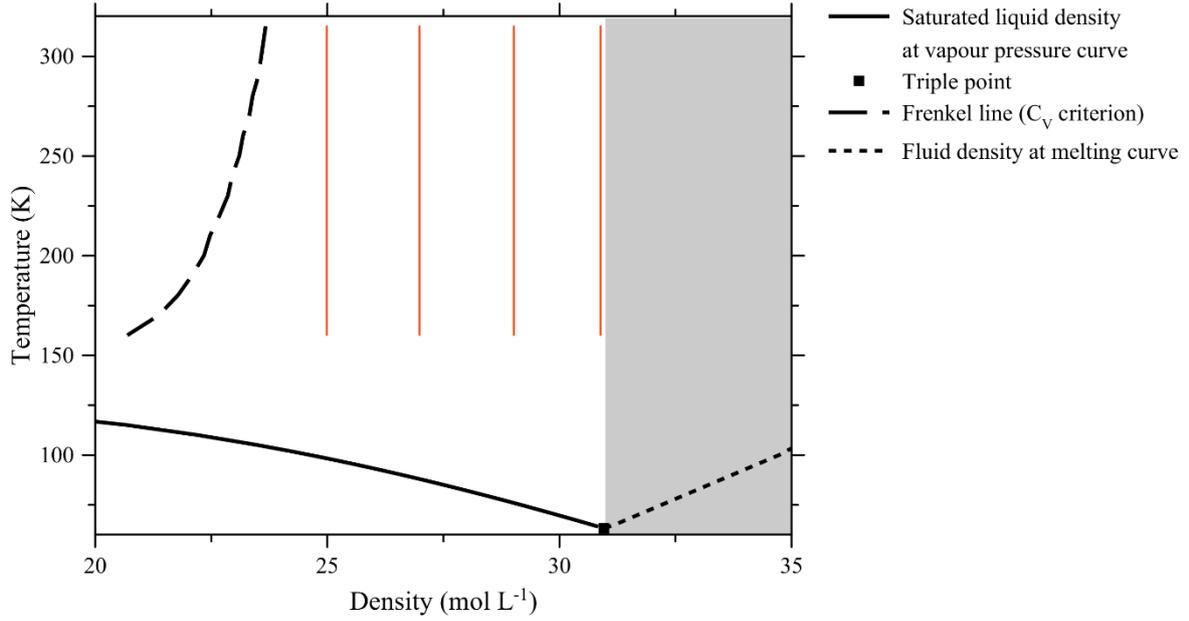

Figure 7. $\rho, T$ diagram of fluid nitrogen showing the first order phase transition lines alongside the Frenkel line. The isochores studied in the present work are marked in red.

When the fluid energy simulation was run for nitrogen along these isochores a good fit was obtained to the experimental $U, C_V$. In this case, $\lambda = 0.5$ was generally preferred. The fitting process only selected $\lambda = 0$ for 25 Mol.L$^{-1}$. A rerun of the calculation with $\lambda = 0.5$ fixed resulted in small changes to $k_D, G_\infty$ and a reduction in the total least squares error in $U$ following the parameter space searching process of about 30%. In figure 8 the discrepancy between the theoretical value of $U$ obtained in the parameter space search at 25 Mol.L$^{-1}$ is shown for $\lambda = 0, 0.5$. As shown, the improvement to the fit obtained by allowing $\lambda = 0$ is limited.

Similarly to Ne, the fitted values of $k_D, G_\infty$ are physically reasonable but do not exhibit the expected trends as a function of increasing density. We therefore ran the calculations with $k_D$ fixed at the average value of $1.17 \times 10^{10}$ m$^{-1}$. Again, this resulted in $G_\infty$ exhibiting a monotonic increase with increasing density as expected. With $k_D$ fixed, $\lambda = 0$ usually resulted from the fitting process.

| Density (Mol.L$^{-1}$) | $\lambda$ | $k_D$ (m$^{-1}$) | $G_\infty$ (GPa) |
|---|---|---|---|
| 25 | 0 (fitted) | $1.32(1) \times 10^{10}$ (fitted) | 1.11(2) (fitted) |
| 25 | 0.5 (fixed) | $1.44(1) \times 10^{10}$ (fitted) | 1.48(2) (fitted) |
| 27 | 0.5 (fitted) | $1.26(1) \times 10^{10}$ (fitted) | 1.66(2) (fitted) |
| 29 | 0.5 (fitted) | $1.13(1) \times 10^{10}$ (fitted) | 1.88(2) (fitted) |
| 30.9 | 0.5 (fitted) | $0.87(1) \times 10^{10}$ (fitted) | 1.42(2) (fitted) |
| 25 | 0 (fitted) | $1.17 \times 10^{10}$ (fixed) | 0.88(1) (fitted) |
| 27 | 0 (fitted) | $1.17 \times 10^{10}$ (fixed) | 1.27(1) (fitted) |
| 29 | 0.0 (fixed) | $1.17 \times 10^{10}$ (fixed) | 1.71(1) (fitted) |
| 30.9 | 0 (fitted) | $1.17 \times 10^{10}$ (fixed) | 2.06(1) (fitted) |

Table 2. Fit parameters for simulations of fluid nitrogen.

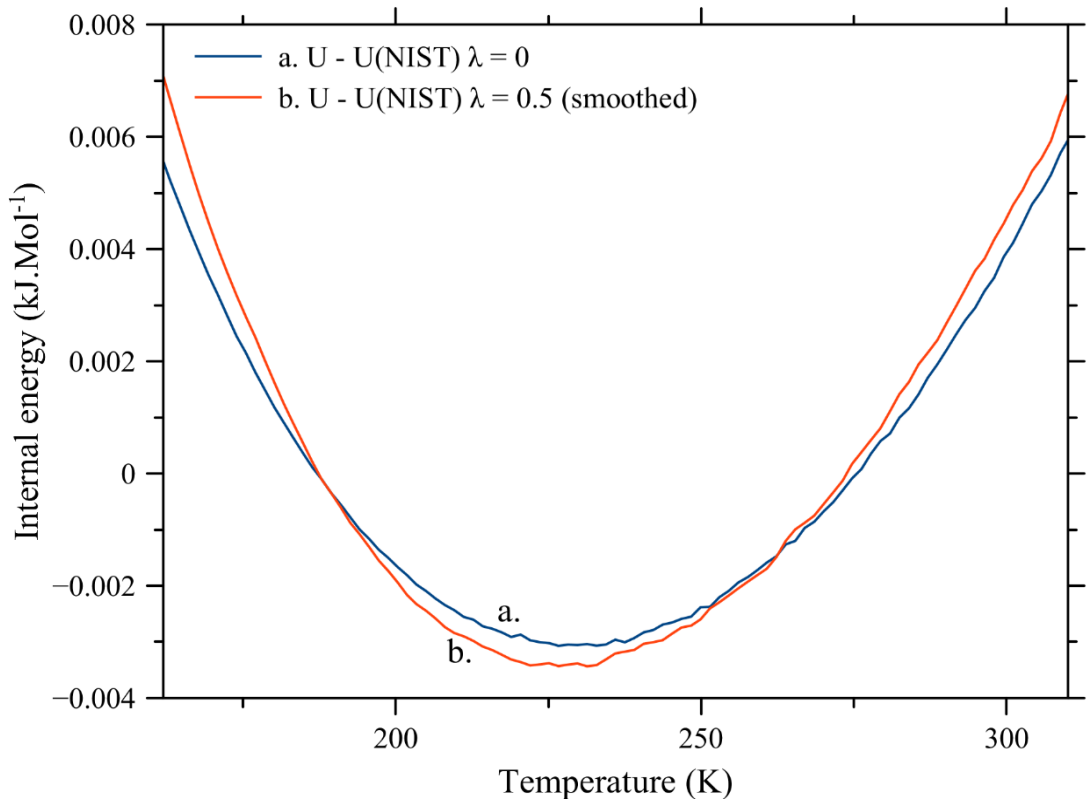

Figure 8. Error in output of the fluid energy program for $U$ following parameter space search for N$_2$ at 25 Mol.L$^{-1}$ for $\lambda = 0, 0.5$.

Throughout the calculations listed in table 2, the adjustments to $k_{prop}$ made during the refinement process to ensure an excellent fit to $C_V$ did not exceed $\pm 2.0\%$ and in most cases did not exceed $\pm 0.7\%$. Similarly to the case of Neon, these changes are within the error in the experimental data, the viscosity output from the fundamental EOS.

### 3.3 Some liquids in the subcritical regime

Most of the $\rho, T$ space occupied by the liquid state is in the supercritical ($T > T_C$) regime in which the lower density limit of the liquid state is the Frenkel line. However, we have also tested our model for fluids in the subcritical regime. In this section we show results for Ar and Kr (Krypton) following isochores beginning from close to the triple point. For Ar we have fitted 100 – 170 K at 35 Mol.L$^{-1}$ (illustrated on figure 2) and for Kr we have fitted 125 – 225 K at 29 Mol.L$^{-1}$, an isochore lying very close to the density of liquid Kr at the triple point. The relatively small temperature range was chosen so that a good fit to the viscosity could be obtained throughout by the VFT law, with no gas-like $\sqrt{T}$ component. After checking this, the fluid energy program was run using the raw viscosity data with no fitting or baseline subtraction whatsoever. In both cases, a good fit to the experimentally observed heat capacity could be obtained with a physically realistic values for $G_\infty$ (2.35(2) GPa for Ar, 3.51(3) GPa for Kr) and $k_D$ (7.3(1) $\times$ 10$^9$ m$^{-1}$ for Ar, 5.9(2) $\times$ 10$^9$ m$^{-1}$ for Kr), with small changes to $k_{prop}$ in the refinement process. The maximum change required for Ar was 2.8% (slightly larger than the uncertainty in the viscosity output from the fundamental EOS of 2%) and the maximum change required for Kr was 0.6%. Unfortunately it has not been possible to get access to the original publication describing the fundamental EOS in use for Kr to obtain the error in the viscosity output, however the changes required to $k_{prop}$ are very small indeed in this case.

The low temperature limit for the Ar results was $1.19 T_T$ and for the Kr results $1.08 T_T$ where $T_T$ is the temperature at the triple point, the lowest temperature at which the liquid state exists. The fluid energy program begins to fail at temperatures extremely close to the triple point. Figure 9 shows the trends in $k_{prop}$ after the parameter space searching, and after refinement, for both Ar and Kr.

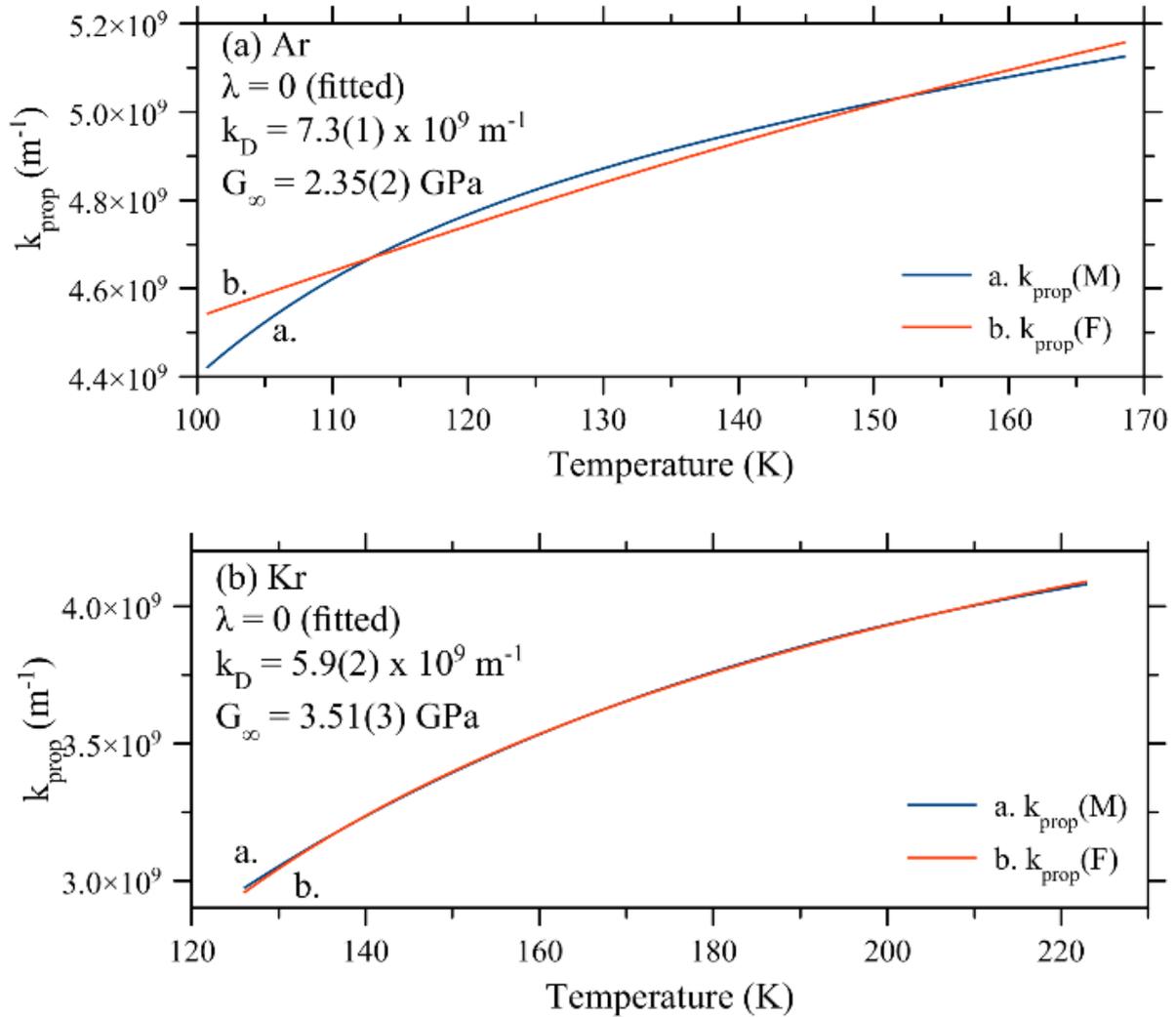

Figure 9. Values of $k_{prop}$ obtained following the parameter space search ($k_{prop}(M)$) and refinement ($k_{prop}(F)$) processes for liquid Ar (a) and Kr (b) in the subcritical regime.

## 4. Discussion and conclusions

The investigations described in sections 2 and 3 constitute the most detailed and rigorous test of the phonon theory of liquid thermodynamics to date. From the studies presented here we conclude that the theory can model the internal energy of fluids over a wide $P,T$ range with sufficient accuracy to model the heat capacity also. The unique behaviour of dense fluid heat capacity (that it decreases upon temperature increase) is successfully accounted for. Where input parameters are required other than the experimental data ($PVT$ EOS, speed of sound, viscosity) we can confidently state that the parameters take best-fit values which (i) are physically reasonable, (ii) are consistent with the available experimental data and (iii) exhibit the expected trends as a function of temperature and density. We can therefore draw some clear conclusions about the input parameters and model.

### 4.1 What values can the input parameters take?

We can make the following specific conclusions about the different input parameters:

1. The Debye wavenumber $k_D$. Within the constraint that $k_{prop} < k_D$ (when this condition is violated our model is expected to fail) the fit quality obtained using the model is not sufficiently well-constrained to allow $k_D$ to be treated as a fitting parameter simultaneously to $G_\infty$. However, the average values of $k_D$ obtained during fitting do not deviate in a systematic manner from the estimates obtained from comparison to the solid state (given in the supplementary material). The approach taken here, of fixing $k_D$ at the average value across the isochores, then rerunning the calculations at this fixed value is therefore appropriate.
2. The infinite-frequency shear modulus $G_\infty$ to be fitted. When the calculations are performed with $k_D$ fixed at an appropriate value, the best-fit value of $G_\infty$ increases slightly upon density increase as expected. In reality the value of $G_\infty$ would vary slightly along the isochore, and this may be the reason why the model can struggle to cover a very large temperature range in a single calculation or to run at temperatures very close to the triple point. However, since experimental measurements of $G_\infty$ across a wide $P, T$ range are not available accounting for any variation in the value of $G_\infty$ along the isochore would necessitate adding at least one more fitting parameter. This is probably not possible without the fit becoming too poorly constrained. The values of $G_\infty$ obtained during fitting are slightly too high. For instance, from MD simulations summarized in ref. 4 we would expect $G_\infty \approx 0.5$ GPa in liquid Ar in the subcritical regime. The values of $G_\infty$ obtained in our calculations on fluid Ar ranged from 2.4 GPa – 4.8 GPa. In addition, the variation as a function of density seems too high for N₂ (see table 2). These discrepancies are reasonable given that the phonon theory of liquid thermodynamics is an approximate treatment (unavoidably so since it is based on the Debye model for solids, which is itself not exact).
3. The anharmonicity parameter λ. The rationale for setting this to 0 or 0.5 was outlined earlier. We observe that the fit quality is only weakly dependent on λ. We therefore conclude that, at least for the fluids studied here, the Grüneisen approximation (that the pressure and temperature dependence of vibrational frequencies can be incorporated into the volume dependence, leading to $\lambda = 0$) could be used throughout for simplicity.

### 4.2 Do the findings provide evidence that the model is correct?

Generally, for a physical model to have any value it must be a model which *could* fail to fit the experimental data if it were wrong – as opposed to a model which automatically fits the data due to the number of adjustable parameters. The phonon theory of liquid thermodynamics unavoidably has a sufficient number of adjustable parameters such that it can always be made to fit the data because very few direct measurements of Frenkel's liquid relaxation time are available, and these measurements do not cover a wide $\rho, T$ range.

However, what we have shown here is that the phonon theory of liquid thermodynamics *could* have failed in a multitude of ways were it not correct. The model could have required $G_\infty$ and/or $k_D$ to exhibit physically unrealistic values or the wrong trend as a function of density in order to fit the data. It did not. On the contrary, the model constrains $G_\infty$ to a narrow range of values of the correct order of magnitude exhibiting the correct trend on increasing density. The model could have required the value of $k_{prop}$ to depart by orders of magnitude from what is expected from an estimation using the Maxwell relaxation time. It did not.

Eq. (5) has been utilized in previous works with the Maxwell relaxation time (a macroscopic property of the fluid) and with the Frenkel relaxation time (defined on an atomic level). From the results presented here, it seems that the model can only be used with the Frenkel relaxation time if it stays close to the Maxwell relaxation time.

It is also appropriate to briefly discuss the application of the model to quantum liquids. The terms "quantum liquid" and "classical liquid" can have different meanings in different contexts. Our model has been applied to liquids which are classical and quantum in the sense of the condition relevant to Eq. (7) (a classical liquid defined as one for which $\hbar\omega \ll k_B T$ for all excitations). However, we have not yet applied it to liquids which are genuinely quantum (the particle de Broglie wavelength is comparable to the dimensions of the box occupied by each particle and the number of quantum states available for the system does not vastly exceed the number of particles)[4]. Only $H_2$ and He at low temperatures meet these criteria.

**4.3 Can the model be used for *a priori* prediction of liquid heat capacity or other properties?**

We observe that tiny changes to $k_{prop}$, often so small that they could be caused by variation of the measured viscosity within experimental error, can change the predicted heat capacity trend on increasing temperature from one which is completely incorrect to one which fits the data closely. Therefore, whilst reproducing known liquid heat capacities is a nice test of the model it would be unwise to use it to predict heat capacity in the absence of experimental data. On the other hand, if $k_{prop}$ and hence the viscosity are very tightly constrained by the observed heat capacity then a reformulation of the model to predict viscosity *a priori* using the observed heat capacity data could be possible in the future.

**Data availability**

The data that support the findings of the study are available from the author upon reasonable request, and the code used to generate the data is available on github: https://github.com/john-e-proctor/fluidenergy

**Supplementary material**

The supplementary material contains all values used for input parameters to the fluid energy program for the results described here, as well as the code and documentation.

**Acknowledgements**

I would like to acknowledge useful discussions with Kostya Trachenko, James Christian and members of the acoustics research group at the University of Salford, as well as advice on coding from David Beynon; even if I did not always heed the advice.